\documentclass[aps,prl,twocolumn,superscriptaddress,showpacs]{revtex4}
\usepackage{graphicx}% Include figure files
\input epsf.tex

\newcommand{\BE}{\begin{equation}}
\newcommand{\EE}{\end{equation}}
\newcommand{\BA}{\begin{eqnarray}}
\newcommand{\EA}{\end{eqnarray}}

\newcommand{\BW}{\begin{widetext}}
\newcommand{\EW}{\end{widetext}}

\begin{document}
\draft

\title{Scattering and Recombination of Two Triplet Excitons in polymer light-emitting diodes}
\author{Y.Meng}
\affiliation{College of Physics, Hebei Normal University,
Shijiazhuang 050016, China} \affiliation{Department of Physics,
Xingtai University, Xingtai, 054001, China} \affiliation{Hebei
Advanced Thin Films Laboratory, Shijiazhuang 050016, China}
\author{B.Di}
\affiliation{College of Physics, Hebei Normal University,
Shijiazhuang 050016, China} \affiliation{Hebei Advanced Thin Films
Laboratory, Shijiazhuang 050016, China}
\author{X.J.Liu}
\affiliation{College of Physics, Hebei Normal University,
Shijiazhuang 050016, China} \affiliation{Hebei Advanced Thin Films
Laboratory, Shijiazhuang 050016, China}
\author{Y.D.Wang}
\affiliation{College of Physics, Hebei Normal University,
Shijiazhuang 050016, China} \affiliation{Hebei Advanced Thin Films
Laboratory, Shijiazhuang 050016, China}
\author{Z. An}
\email[Corresponding author. Email: ] {zan@mail.hebtu.edu.cn}
\affiliation{College of Physics, Hebei Normal University,
Shijiazhuang 050016, China} \affiliation{Hebei Advanced Thin Films
Laboratory, Shijiazhuang 050016, China}
\date{\today}

\begin{abstract}

The scattering and recombination processes between two triplet
excitons in conjugated polymers are investigated by using a
nonadiabatic evolution method, based on an extended
Su-Schrieffer-Heeger model including interchain interactions. Due
to the interchain coupling, the electron and/or hole in the two
triplet excitons can exchange. The results show that the
recombination induces the formation of singlet excitons, excited
polarons and biexcitons. Moreover, we also find the yields of
these products, which can contribute to the emission, increase
with the interchain coupling strength, in good agreement with
results from experiments.
\end{abstract}


\maketitle

In recent years polymer semiconductors have made dramatic advances
in light-emitting device performance, e.g., polymer light-emitting
diodes (PLEDs) \cite{Shinar}. In these devices, following charge
carrier recombination, singlet and triplet excitons are
subsequently created. Compared to singlet excitons, which can
decay radiatively and contribute to the emission, triplet excitons
are nonluminescent. Although the precise ratio of singlets to
triplets is currently debated, it has been established that
triplet excitons constitute the dominant species in PLEDs
\cite{Shuai,Beljonne,Barford,Tandon,Bittner}. Therefore, research
into the dynamics of triplet excitons in conjugated polymers is
urgently required.

Due to their long lifetime, a large number of the triplet excitons
will accumulate in working PLEDs and have a significant
probability of interacting and recombine with other particles. In
our previous work \cite{Meng}, the recombining of a triplet
exciton and a polaron was carefully investigated. Our results
showed that the nonemissive triplet exciton is converted to
emissive products, i.e., a singlet exciton and a excited polaron
state. Therefore, the quantum efficiency of PLEDs can be enhanced
by polaron-exciton recombination. In addition to this process,
there is also another important case affecting the working
characteristics of PLEDs, namely, triplet-triplet annihilation
(TTA) \cite{Sinha, Iwasaki, Baldo}. These works have demonstrated
that in PLEDs delayed fluorescence (DF) originates from the
production of singlet excitons through the TTA process.
Furthermore, Partee et al. \cite {Partee} found that the DF
intensity in films is higher than that in solutions. Accordingly,
the triplet lifetimes are seen to decrease from solutions to
films. In a recently performed experiment, Ribierre et al.
\cite{Ribierre} studied the effect of intermolecular spacing on
the physics of exciton diffusion and light emission. They found
the rate of TTA which is correlated to exciton diffusion,
decreases with the intermolecular distance. These two works
\cite{Partee, Ribierre} showed that both T-T annihilation by
intermolecular recombination and the total emission are higher in
the presence of strong interchain interactions. Although some
possible mechanisms for TTA processes have been proposed, no
detailed theoretical investigations of how triplet exciton
recombination affects quantum efficiency have been reported.

The goal of the present work is to investigate the recombination
of two triplet excitons on two coupled chains and to address the
branching ratio of the products. We use a nonadiabatic evolution
method at the unrestricted Hartree-Fock level within an extended
Su-Schrieffer-Heeger model including interchain interactions, with
the extended Hubbard model for electron-electron interactions. Due
to the interchain coupling, the electron and/or hole in the two
triplet excitons can transfer between the two chains. As a result,
the two triplet excitons can recombine to form singlet excitons,
as well as excited polarons and biexciton states. These products
are all luminescent due to radiative decay, which can enhance the
efficiency of PLEDs. Furthermore, when compared with the results
of experiments \cite{Sinha, Iwasaki, Baldo, Partee}, which
indicate that the DF derives from singlet excitons formed in TTA
process, our results show that the primary source of DF is excited
polarons.

The one chain model Hamiltonian we use is the Su-Schrieffer-Heeger
(SSH) model \cite{Su} with a Brazoskii-Kirova-type
symmetry-breaking term \cite{Brazovskii}, including the effect of
the intrachain electron-electron (e-e) interaction \cite{Hubbard},
\begin{eqnarray}
H &=& -\sum_{j,n,s} t_n \left( c_{j,n,s}^{\dagger
}c_{j,n+1,s}+c_{j,n+1,s}^{\dagger }c_{j,n,s}\right) \nonumber \\
&&+\frac K2\sum_{j,n}(u_{j,n+1}-u_{j,n})^2+\frac
M2\sum_{j,n}\stackrel{.}{u}_{j,n}^2 \nonumber \\
&&+U\sum_{j,n} c_{j,n,\uparrow}^{\dagger}
c_{j,n,\downarrow}^{\dagger}c_{j,n,\downarrow}c_{j,n,\uparrow} \nonumber \\
&&+V\sum_{j,n,s,s^{\prime}}
{c_{j,n,s}^{\dagger}c_{j,n+1,s^{\prime}}^{\dagger}c_{j,n+1,s^{\prime}}c_{j,n,s}},
\label{eq1}
\end{eqnarray}
Here, $j=1,2$ is the chain index. The quantity $t_n$ is given by
$t_n=t_0-\alpha (u_{j,n+1}-u_{j,n})+(-1)^nt_e$ with $t_0$ being
the transfer integral of $\pi $-electrons in a regular lattice,
$\alpha $ the electron-lattice coupling constant, and $u_{j,n}$
the lattice displacement of the atom at the $n$-th site along the
$j$-th chain from its equidistant position. The quantity $t_e$ is
introduced to lift the ground-state degeneracy for nondegenerate
polymers. The operator $c_{j,n,s}^{\dagger}(c_{j,n,s})$ creates
(annihilates) a $\pi $-electron at the $n$-th site with spin $s$
along the $j$-th chain; $K$ is the elastic constant due to the
$\sigma $ bonds, and $M$ is the mass of a CH group. $U$ gives the
strength of the on-site Coulomb interactions and $V$ gives the
strength of the nearest-neighbor interactions.

The two chains are coupled by the interchain interactions
$H^{\prime}$ \cite{Meng1},
\begin{eqnarray}
H^{\prime}&=&-\sum_{n,s} t_\bot \left(
c_{1,n,s}^{\dagger}c_{2,n,s}+c_{2,n,s}^{\dagger}c_{1,n,s}\right)\nonumber \\
&&+\sum_{n,s,s^{\prime}} V_\bot
c_{1,n,s}^{\dagger}c_{2,n,s^{\prime}}^{\dagger}c_{2,n,s^{\prime}}c_{1,n,s},\label{eq2}
\end{eqnarray}
where $ t_{\bot} $ stands for the transfer integral between sites
labelled by the same index $n$ on the two chains and $V_\bot$ is
the e-e interaction term. In the following calculations $V_\bot$
is fixed at $0.1\rm{eV}$.

The temporal evolution of the lattice is determined by the
equation of motion for the atomic displacements,
\begin{eqnarray}
M\stackrel{..}{u}_{j,n}& =& -K\left(
2u_{j,n}(t)-u_{j,n+1}(t)-u_{j,n-1}(t)\right) \nonumber \\
&&+2\alpha \sum_s \left[ \rho _{jn,jn+1}^s(t) -\rho
_{jn,jn-1}^s(t) \right], \label{eq3}
\end{eqnarray}
where the density matrix $\rho $ is defined as
\begin{equation}
\rho _{jn,j^{\prime }n^{\prime }}^s\left( t\right) =\sum_k\Phi
_{j,n,k}^s\left(t\right) f_{k,s}\Phi _{j^{\prime
},n^{\prime},k}^{s*}(t), \label{eq4}
\end{equation}
with $f_{k,s}$ being the time-independent distribution function as
determined by the initial electron occupation. The electronic wave
functions $\Phi _{j,n,k}^s\left(t\right)$ are the solutions of the
time-dependent Schr\"{o}dinger equation
\begin{equation}
i\hbar \stackrel{.}{\Phi }_{j,n,k}^s (t)
=\sum_{j^{\prime},n^{\prime}}h_{jn,j^{\prime}n^{\prime}}^s
(t){\Phi_{j^{\prime},n^{\prime},k}^s (t)}. \label{eq5}
\end{equation}
The coupled differential equations (\ref{eq3}) and (\ref{eq5}) can
be solved with a Runge-Kutta method of order 8 with step-size
control \cite{Brankin}. The parameters used here are,
$t_0=2.5\rm{eV}$, $\alpha =4.1\rm{eV/\AA}$, $t_e=0.05\rm{eV}$,
$K=21\rm{eV/\AA ^2}$, and $M=1349.14\rm{eV\cdot fs^2/\AA ^2}$.

In our numerical calculations, two 50-site parallel polymer chains
opposite each other are considered: the CH-units are labeled as
1-50 on chain 1 and 51-100 on chain 2. We start with the
stationary solution to the system after two triplet excitons are
introduced, respectively, into chains 1 and 2. The initial bond
configuration and electronic structure can be obtained by
minimizing the total static energy of the two-chain system in the
absence of the interchain interactions.

Figure 1 (a) shows the initial staggered bond order parameters
${\delta}_n\equiv \left( -1\right) ^n\left(u_{n-1} + u_{n+1} -
2u_n\right)/4 $ and a schematic diagram of the energy levels for
the two triplet excitons. In polymers, the exciton is a composite
particle consisting of an electron and a hole bound together by a
lattice deformation, due to the strong electron-lattice
interactions. With the two deformations, four localized energy
levels emerge in the energy gap. We consider the two excitons to
have opposite spins (with $S_Z=+1$ and $S_Z=-1$). Obviously, there
are two pairs of degenerate energy levels in the absence of
interchain coupling. The corresponding wave functions are
presented in Fig. 1 (b) [the 49th and 51st energy levels are
localized on chain 1, and the 50th and 52nd energy levels are
localized on the chain 2].

\begin{figure}
\epsfxsize=3in\epsffile{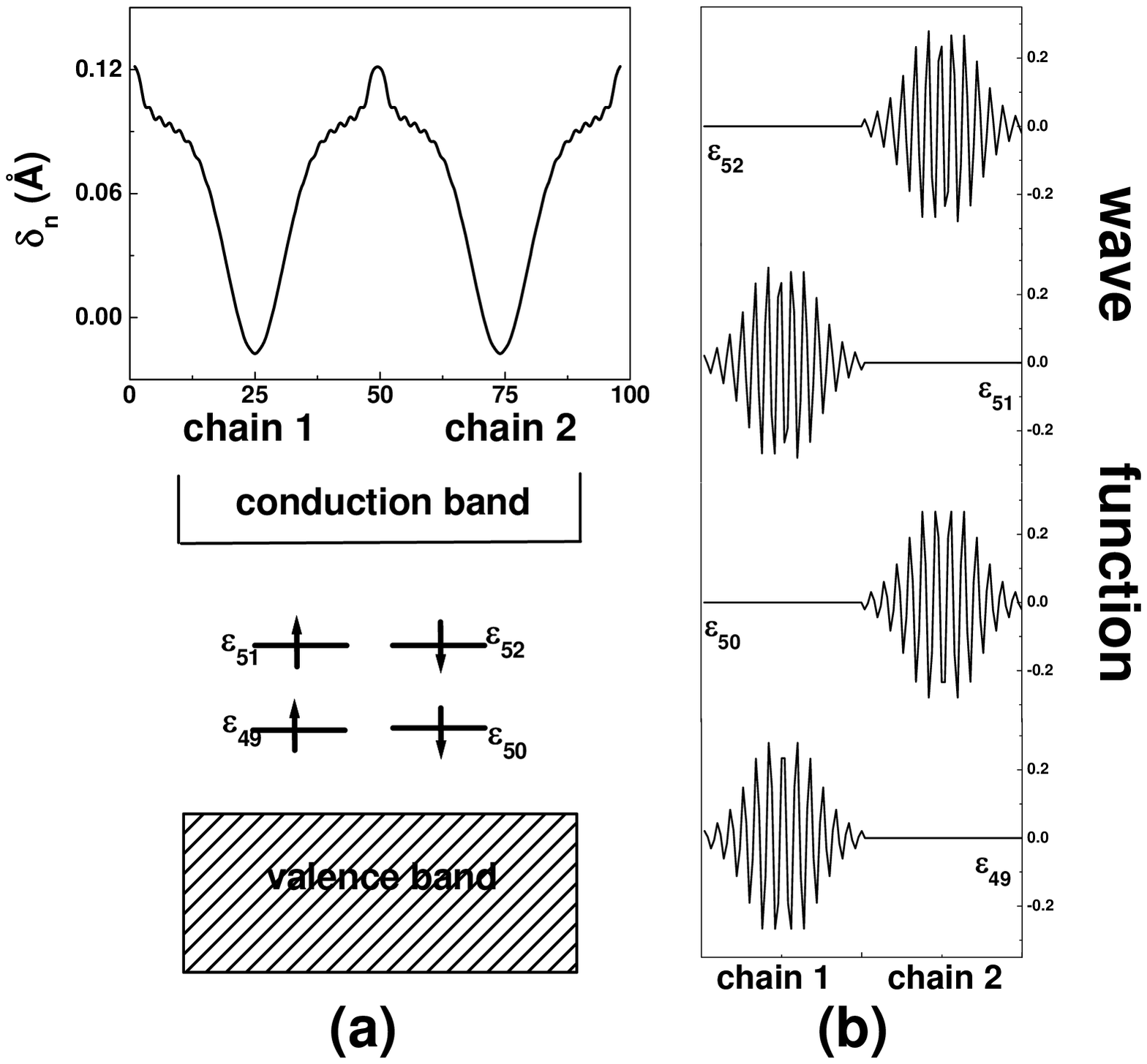} \caption{(a) The
staggered bond order parameters ${\delta}_n$ and schematic diagram
of energy levels for the two triplet excitons; (b) The wave
functions of the gap-state (from 49th energy level to 52th energy
level).} \label{fig1}
\end{figure}

Then, turning on the interchain interactions, we simulate the
charge transfer process and the dynamic relaxation of the
recombination, with special attention given to the branching ratio
of various species formed in the steady state. In order to avoid
abrupt changes, the interchain coupling strength is smoothly
increased over a period of 400fs, and then held at a constant
value. As expected, due to the interchain coupling, the electron
and /or hole in the two excitons will exchange between the two
chains. That is to say, a portion of the spin-up electron and/or
hole localized initially on chain 1 transfers to chain 2, and the
spin-down electron and/or hole localized initially on chain 2
transfers partially to chain 1. Because the total charges on the
two chains are equal initially, the effects of the interchain
interactions only redistribute the electron and/or hole with
different spin on the two chains and do not change the quantity of
charge on each chain. Correspondingly, the amplitudes of the two
lattice distortions for excitons are unchanged in the dynamical
evolution processes. When the interchain coupling strength
eventually increases to its final value, the charge transfer also
reaches a maximum. At this point, the system approaches a
dynamically stable state.

We now analyze the products of the recombination and determine the
branching ratios for the formation of various species in the
steady state. The approximate eigenstates $\chi_1$ (localized on
one chain) and $\chi_2$ (localized on the other chain) can be
separately obtained by ignoring the interchain coupling. It is
found that such a structure is a mixture of sixteen possible
states, i.e., an excited polaron localized on chain 1 and a
polaron localized on chain 2, a polaron localized on chain 1 and
an excited polaron localized on chain 2, etc. For simplicity, we
divide these states into five species: (a) a triplet exciton on
each chain, (b) a singlet exciton on each chain, (c) an excited
polaron and a polaron, (d) a negative bipolaron and a positive
bipolaron, and (e) a biexciton and a ground state, see Fig. 2
[only the gap-states, ${\chi}^u_{1}$, ${\chi}^d_{1}$ for the
particle localized on one chain, and ${\chi}^u_{2}$,
${\chi}^d_{2}$ for the particle localized on another chain, are
shown].

\begin{figure}
\epsfxsize=2.5in\epsffile{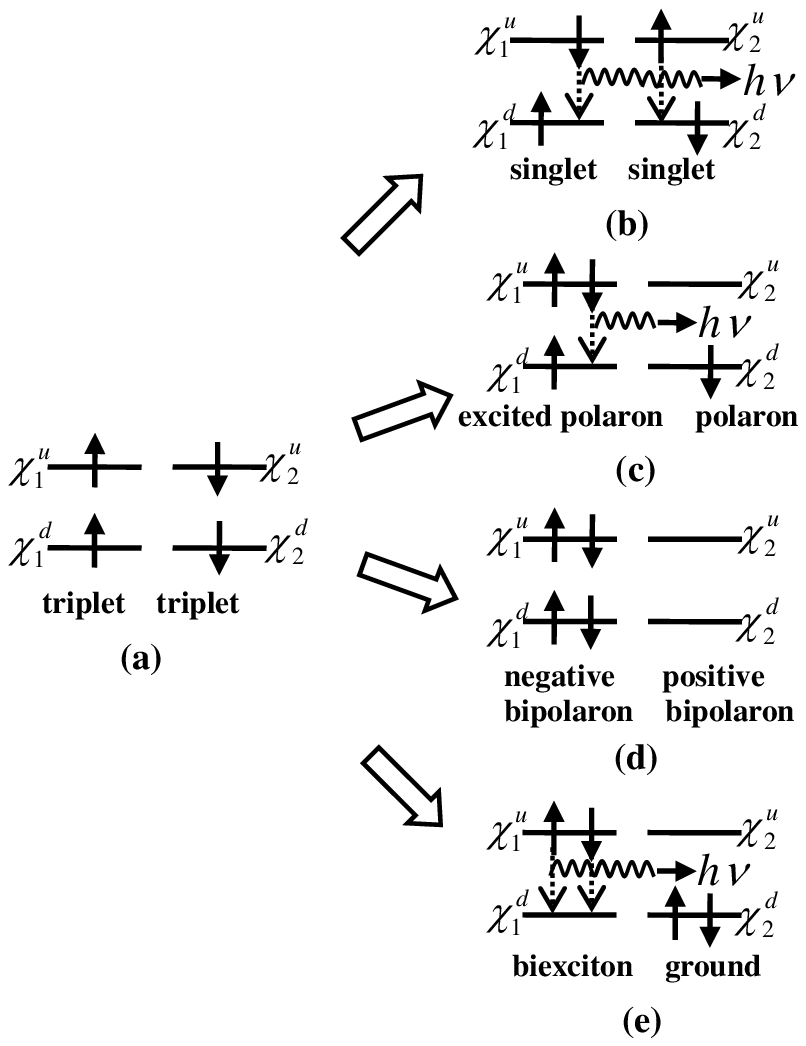} \caption{ The possible
species of the recombination products of two triplet excitons.}
\label{fig2}
\end{figure}

\begin{figure*}
\epsfxsize=5.0in\epsffile{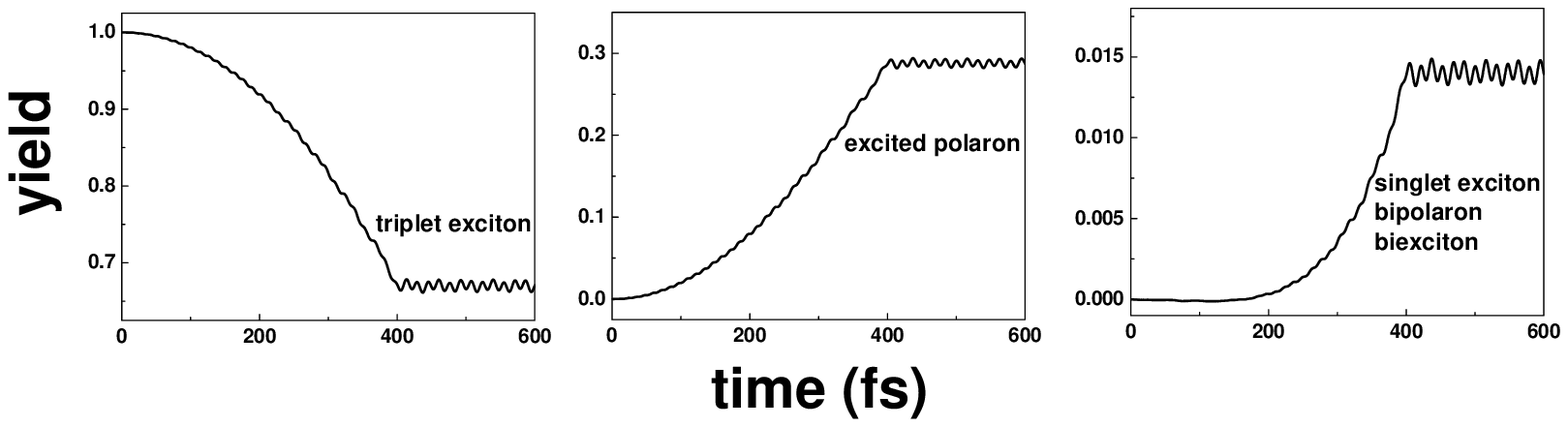} \caption{The yields of
five species vary with time in the case of $t_\bot=0.1\rm eV$.}
\label{fig3}
\end{figure*}

In order to ascertain the yields of these states, after each
evolution step we can project the evolution wave functions $|\Phi
\left(t\right)\rangle$ onto the space of the eigenstates of the
system. The relative yield $I_K (t)$ for a given eigenstate
$|\Psi_K\rangle$ is then obtained from
\begin{equation}
I_K (t)=|\langle\Psi_K|\Phi \left(t\right)\rangle|^2. \label{eq6}
\end{equation}
The evolved wave function of the whole system $|\Phi
\left(t\right)\rangle$ can be constructed using the single
electron evolution wave function $|\Phi
_{j,n,k}^s\left(t\right)\rangle$ written as a Slater determinant.
$|\Psi_K\rangle$ denotes the states of products shown in Fig. 2
(a)-(e). As an example, Fig. 3 shows how the yields of five
species vary with time for the case $t_\bot=0.1\rm eV$. It may be
seen that the yield of triplet excitons decreases, and
simultaneously, the yields of the other four states increase as
the interchain interactions are slowly increased. That is to say,
the initial triplet excitons, due to interchain charge transfer,
are transformed little by little into the other four states. After
the interchain coupling strength reaches a constant value (at
400fs), the yields tend to be steady with only slight
oscillations. In detail, the yields of the triplet exciton, the
singlet exciton, the excited polaron, the bipolaron, and the
biexciton are about 66.8\%, 1.4\%, 28.2\%, 1.4\%, and 1.4\%,
respectively.

It should be stressed that the excited polaron is luminescent as
is the singlet exciton because they have comparable transition
dipole moments (about 30{e\AA} for the singlet exciton and
54{e\AA} for the excited polaron). Furthermore, the biexciton
state, in which two electrons and two holes are trapped together
by a lattice distortion, can decay to the ground state by emitting
two photons. Therefore, our results indicate that two nonemissive
triplet excitons can be transformed into the emissive singlet
exciton, the emissive excited polaron, and the emissive biexciton.
Based on simple spin statistics arguments, the yield of the
triplet exciton should be 75\% and that of the singlet exciton
should be 25\% in PLEDs. When the recombination of two triplet
excitons is taken into account, as stated above, for the case
$t_\bot=0.1\rm eV$, two triplet excitons have a probability of
about 31\% of being converted into emissive species (28.2\% to the
excited polaron, 1.4\% to the singlet exciton, and 1.4\% to the
biexciton, see fig. 3). Other processes also play a role. For
example, in the recombination of the two excitons with parallel
spin ($S_Z=+1$ or $S_Z=-1$), new products do not form. In
contrast, for the recombination of the two excitons with $S_Z=+1$
and $S_Z=0$, or $S_Z=-1$ and $S_Z=0$, or both $S_Z=0$, the
emissive excited polaron which is the primary product can be
formed by charge exchange in the two chains. To summarize the
above, in most processes where two triplet excitons recombine
(with $S_Z=+1, 0,$ and $-1$), emissive products can be formed
through interchain interactions. Therefore, the quantum efficiency
of light-emitting diodes is enhanced, and the primary source of
the increased efficiency is the excited polaron. Consequently, the
experimental fact that the quantum efficiency exceeds 25\%
\cite{Cao, Ho} can be understood on the basis of recombination of
two excitons.

With increasing interchain coupling strength, charge transfer
between the two chains becomes easier, i.e., the transformation of
various electron/hole states becomes easier. As a result, the
products of the triplet-triplet recombination, i.e., the excited
polaron, the singlet exciton, the bipolaron, and the biexciton
increase with increasing interchain coupling strength, while the
yield of the triplet exciton decreases, see Fig. 4. This implies
that the interchain interaction greatly increases the formation of
the emissive excited polarons, singlet excitons and biexcitons,
and thus the quantum efficiency of devices. Our calculated results
are in good agreement with the experimental observations
\cite{Partee, Ribierre}. In ref 11 the delayed fluorescent of
poly(p-phenylene vinylene) (PPV) and poly(p-phenylene ethynylene)
(PPE) derivative solids and solutions was described. The results
show that triplet-triplet bimolecular recombination induces the
DF. Furthermore, the DF intensity in films is higher than that in
solutions and the triplet lifetimes decrease from solutions to
films. Reference 12 described the dynamics of triplet exciton
quenching in phosphorescent dendrimers, in which the spacing of
the emitting core was controllable. They found that as the
distance between the cores increased, the rate of TTA decreased.
These results showed that both T-T annihilation by intermolecular
recombination and the total emission are higher in strongly
interchain interactions. Therefore, our calculated results are
indeed consistent with those found experimentally.

\begin{figure}
\epsfxsize=2.5in\epsffile{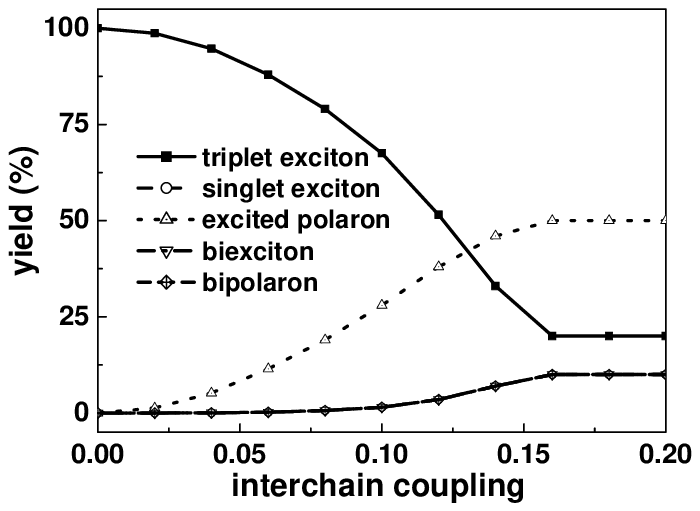} \caption{The five
species included in the final state as functions of the interchain
coupling strength.} \label{fig4}
\end{figure}

\begin{figure*}
\epsfxsize=5.0in\epsffile{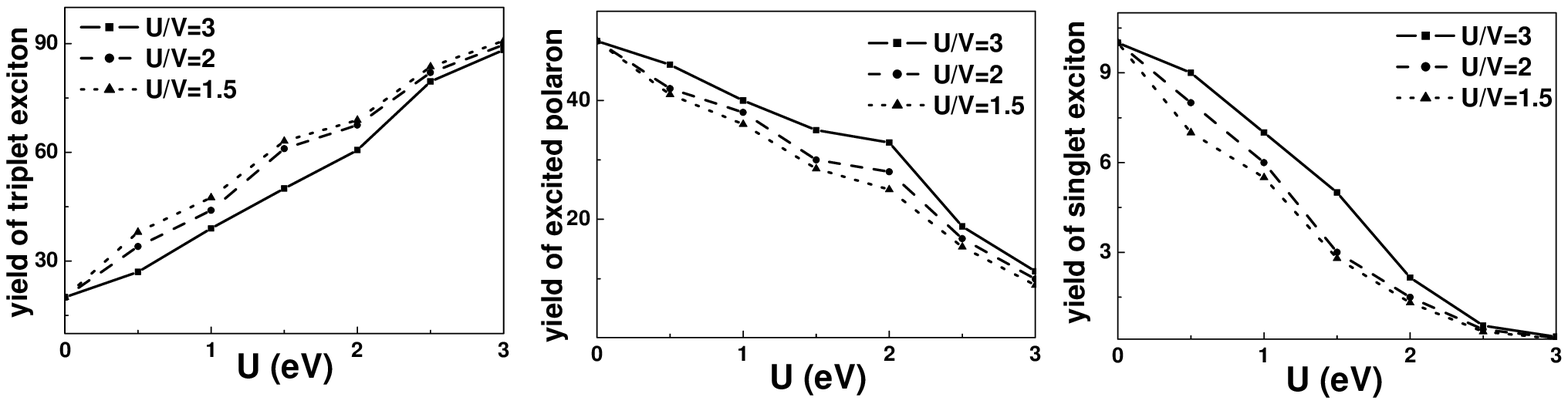} \caption{The yields of
the triplet exciton, singlet exciton and excited polaron with
different e-e interactions.} \label{fig5}
\end{figure*}

Finally, the effects of the e-e interactions on recombination are
discussed. Depicted in Fig. 5 are the yields of the singlet
exciton, which is equal to that of biexciton, and the excited
polaron as functions of the on-site Hubbard energy U for different
nearest neighbor Coulomb interaction strengths V with
$t_\bot=0.1\rm eV$. The value of U is allowed to range from 0 to
3.0 eV, and the relationships between U and V are U=3.0V (solid
line), U=2.0V (dashed line), U=1.5V (dotted line). It is found
that the emissive singlet exciton, biexciton and the excited
polaron are suppressed by U and V. The reason is that the binding
energy of the electron and hole in a self-trapping excited state
increases with the e-e interactions, making charge transfer
between the two chains more difficult. Therefore, strong e-e
interactions are unfavorable for the formation of the emissive
species.

In conclusion, we have simulated the two triplet exciton
recombination processes using a nonadiabatic evolution method. We
have identified four types of products, of which the singlet
exciton, the excited polaron, and the biexciton can decay
radiatively. Our results indicate that the quantum efficiency of
polymer light-emitting diodes is enhanced and can exceed 25\% (the
simple statistical limit) by the recombination of two triplet
excitons. Furthermore, interchain interaction greatly increases
the formation of emissive species, in good agreement with
experimental observations. The effects of the e-e interactions on
the recombination processes are also discussed.

This work was partially supported by the National Natural Science
Foundation of China (Grant Nos. 11074064 and 10574037), Program
for New Century Excellent Talents in University of Ministry of
Education of China (Grant No. NCET-05-0262), Hebei Provincial
Outstanding Youth Science Fund (Grant No. A2009001512), Key
Project of Ministry of Education of China (Grant No.210021), and
Natural Science Fund of Hebei Province, China (Grant No.
A2010000357).

\end{document}